\documentclass[conference,11pt]{IEEEtran}

\usepackage[utf8]{inputenc}
\usepackage{float}

\newcommand{\eat}[1]{}

\usepackage{xcolor}

\usepackage{graphicx}
\begin{document}
\title{Spoofing Attacks Against Vehicular FMCW Radar}

\author{\IEEEauthorblockN{Rony Komissarov}
\IEEEauthorblockA{\textit{School of Electrical Engineering} \\
\textit{Tel Aviv University}\\
Ramat Aviv, 69978, Israel \\
ronykom@gmail.com}
\and
\IEEEauthorblockN{Avishai Wool}
\IEEEauthorblockA{\textit{School of Electrical Engineering} \\
\textit{Tel Aviv University}\\
Ramat Aviv, 69978, Israel \\
yash@eng.tau.ac.il}

	}

\maketitle

\begin{abstract}
The safety and security of the passengers in vehicles in the face of cyber attacks is a key element in the automotive industry, especially with the emergence of the Advanced Driver Assistance Systems (ADAS) and the vast improvement in Autonomous Vehicles (AVs). Such platforms use various sensors, including cameras, LiDAR and mmWave radar. These sensors themselves may present a potential security hazard if exploited by an attacker.

In this paper we propose a system to attack an automotive FMCW mmWave radar, that uses fast chirp modulation. Using a single rogue radar, our attack system is capable of spoofing the distance and velocity measured by the victim vehicle simultaneously, presenting phantom measurements coherent with the laws of physics governing vehicle motion. The attacking radar controls the delay in order to spoof its distance, and uses phase compensation and control in order to spoof its velocity. After developing the attack theory, we demonstrate the spoofing attack by building a proof-of-concept hardware-based system, using a Software Defined Radio. We successfully demonstrate two real world scenarios in which the victim radar is spoofed to detect either a phantom emergency stop or a phantom acceleration, while measuring coherent range and velocity.  We also discuss several countermeasures to the attack, in order to propose ways to mitigate the described attack.
\end{abstract}

\section{Introduction}
\subsection{Background}
The safety and security of the passengers in vehicles in the face of cyber attacks is a key element in the automotive industry, especially with the emergence of the Advanced Driver Assistance Systems (ADAS) and the vast improvement in Autonomous Vehicles (AVs)~\cite{ziebinski2016survey}. The main sensors available on modern self-driving cars are Ultrasonic sensors, mmWave (Millimeter Wave) radars, cameras, and LiDAR~\cite{wang2019multi,fung2017sensor}. These sensors are used for sensing the physical environment and for safety-critical decisions, such as collision avoidance and intersection management. Adversarial sensor attacks manipulate the input signals to the sensors in order to produce incorrect environment views with the goal of causing unsafe actions. These attacks can be organized into two main categories: Spoofing and Jamming~\cite{yan2016can}. While the latter attack can be easily detected, mitigating a spoofing attack usually requires more intelligent countermeasures~\cite{kapoor2018detecting}. Spoofing makes it very difficult for the sensor system to recognize that it is under attack, as it provides the victim sensor with seemingly legitimate but actually false data. In this work we focus on spoofing a vehicle's mmWave radar.

\subsection{Related Work}
Various sensors of autonomous vehicle are vulnerable to spoofing attacks. Prior work mainly focused on the study of security in perception sensors, primarily the LiDAR and camera. Physical world camera attacks were demonstrated by fooling Deep Neural Networks (DNNs), making objects mislabeled or ignored~\cite{song2018physical,eykholt2018robust}. For example, in~\cite{eykholt2018robust}, the researchers caused a ``STOP'' sign disappear in the eyes of the detector, by adding adversarial stickers onto the sign. Attacks against LiDAR-based perception systems in AVs appear in~\cite{cao2019adversarial,changalvala2019lidar}. The authors found that simply spoofing the LiDAR using common methods is not sufficient due to the use of machine learning object detection models. To mount an attack they leveraged an optimization method to generate adversarial examples to fool a real world working LiDAR.

Some work has also been done to attack and spoof ranging systems, such as Radar and Ultrasonic sensors~\cite{thing2016autonomous}. Due to the fact that the mmWave radar is crucial in bad weather conditions, an attack to fool its measured distance is a critical physical security threat~\cite{goodin2019predicting,ranganathan2012physical}.
Specifically, a distance-only spoofing attack~\cite{miura2019low,nashimoto2021low} was presented on a mmWave Frequency Modulated Continuous Wave (FMCW) radar. The system used a weak Arduino platform against an FMCW radar which used the less-common triangular waveform. The attackers synchronised to the victim via half-chirp modulation, and were able to control the delay measured by a victim radar to spoof the range. However, beyond measuring range, FMCW radars are typically used to measure velocity, which they can achieve with high precision~\cite{fung2017sensor}. The attack method in~\cite{miura2019low} does not spoof the velocity  measured by the victim at all.

Recently, the researches in~\cite{sun2020control} constructed several attack scenarios in order to spoof a mmWave radar of a Lincoln MKZ-based AV testbed, using a Software Defined Radio (SDR) transceiver system from National Instruments. The attack strategies involved synchronised attack radars on both sides of the road, using a complex setup. Furthermore, the method the attackers used to spoof velocity measurements was by spoofing the distance at subsequent time intervals. Since modern mmWave radars have the ability to measure velocity independently from distance, even a simple intra-radar sensor fusion approach can mitigate the attacks described, since the measured range and velocity are not coherent~\cite{jagielski2018threat}.       

\subsection{Our Contribution}
In this work we demonstrate an adversarial radar based on a SDR, which can simultaneously manipulate the range and velocity measured by a victim radar used by an ADAS or AV platform.

The adversarial radar we designed utilizes the advantages of a complex baseband FMCW radar: beyond controlling the delay to spoof the range, we are able to manipulate the signal phase received at the victim radar's Rx antenna, to spoof the velocity.

Unlike~\cite{miura2019low,nashimoto2021low}, which focused on radars using a triangular chirp waveform, our system addresses the modern saw-tooth waveform (so called ``fast chirps''), as this chirp waveform is frequently found in contemporary FMCW radars in the AV industry~\cite{lutz2014fast,tong2015fast}. And unlike~\cite{sun2020control}, we only require a single rogue radar within a vehicle in front of the victim.

After developing the attack theory, we demonstrate the spoofing attack by building a proof-of-concept hardware-based system, using a bladeRF Software Defined Radio. Using the ability to manipulate the velocity and the range measured by the victim radar, we demonstrate two realistic automotive attack scenarios: spoofing a phantom emergency break, and spoofing a phantom acceleration. In both cases, the range and velocity measured by the victim radar are coherent and fit the laws of physics governing vehicular motion.

\section{The FMCW Radar}

Generally, Radars emit electromagnetic waves and receive the reflection to measure the time of flight. 
Specifically, in FMCW radar solutions, the transmitted signal is a linear frequency modulated continuous wave (FMCW) chirp~\cite{mitomo201077}.  
FMCW radars are common in a variety of industries and applications, such as naval navigation radars, smart ammunition sensors, industrial radars, and in particular automotive radars~\cite{stove1992linear,lin2016design,komissarov2019partially}.
 
 Since the frequency component of the chirp is non constant in time, it is usually described as a frequency over time spectrogram, as described in Figure~\ref{fig:Range Measurement}(a).
 The description of the chirp development in time is given in Figure~\ref{fig:chirp tx and rx}(a).

Modern systems use a saw-tooth waveform chirp with slope $S = \frac{df}{dt}$ and  bandwidth parameters depending on the distance and velocity of the object of interest~\cite{winkler2007range,tong2015fast}. The state of the art FMCW radars integrated in commercial vehicles use the 76-81 GHz frequency band~\cite{gao2019experiments}. 
Typical FMCW Radar implementations include a Voltage Controlled Oscillator (VCO) which is used to create the necessary chirp waveform. The output of the VCO is amplified by the Power Amplifier (PA), and then transmitted through the Antenna. 
In order to improve the precision of the physical measurements, a series of N chirps (a frame) is averaged. The number of chirps $N$ is usually in the range of 50 to 1000~\cite{lutz2014fast}.

\subsubsection{Measuring Range Using FMCW Radar}
\begin{figure}[t]
\centering
\includegraphics[scale=0.4]{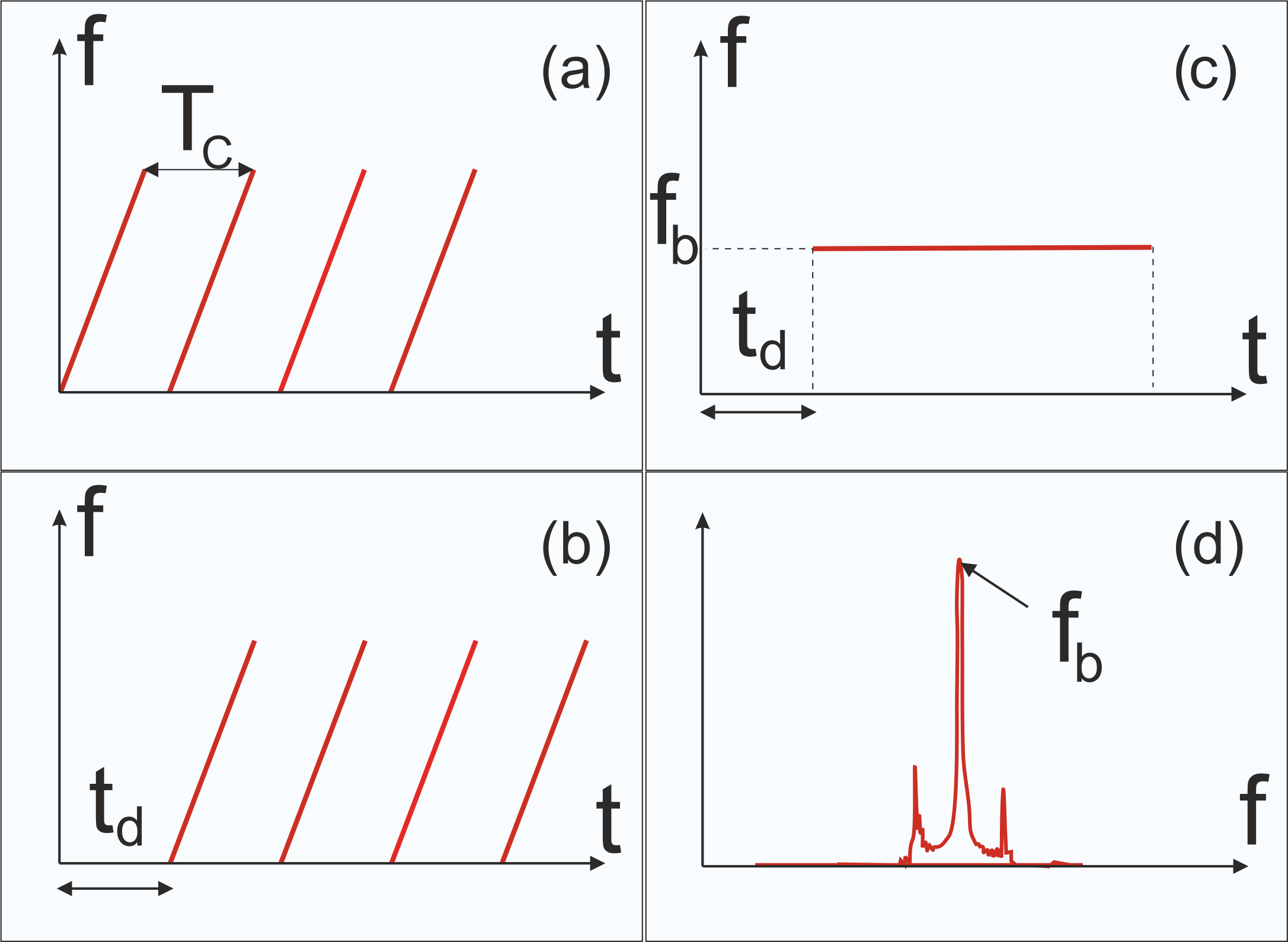}
\caption{Range measurement using FMCW radar: (a) Spectrogram of a frame of saw-tooth chirps with a chirp duration $T_C$. (b) The spectrogram reflected from the object of interest with returned chirps showing a delay of $t_d$. (c) The spectrogram of the IF signal at the receiver, with a beat frequency of $f_b$. (d) Demonstration of range-FFT in order to find the range corresponding to the beat frequency. }
\label{fig:Range Measurement}
\end{figure}

The transmitted FMCW signal is reflected by the object of interest, and gets back to the Rx antenna of the radar after a total delay time $t_d$, as seen in Figure~\ref{fig:Range Measurement}(b). The received chirp is then mixed with the currently transmitted signal. The product of the mixing process is called an IF (Intermediate Frequency) signal, and it exhibits a single beat frequency of $f_{b}$ (Figures~\ref{fig:Range Measurement}(c),(d)). Finally, the IF signal is sampled by the ADC and further processed by a DSP or MCU. The range $d$ of the object of interest is related to the beat frequency $f_{b}$ by the following equation:
\[d = \frac{c\cdot f_{b}}{2S}\]
where S is the slope of the Tx chirp and $c$ is the speed of light.

There are several ways to extract the beat frequency of the IF signal. A common way is to use a fast Fourier transform (Range-FFT), to convert the time domain signal into a spectrum with a peak at $f_{b}$, as depicted in Figure~\ref{fig:Range Measurement}(d). Typical IF frequencies are in the range of a few hundred kHz for objects within 100m. Thus, there is no requirement for a relatively high speed ADC.

\subsubsection{Measuring Velocity Using FMCW Radar}
\begin{figure}[t]
\centering
\includegraphics[scale=0.105]{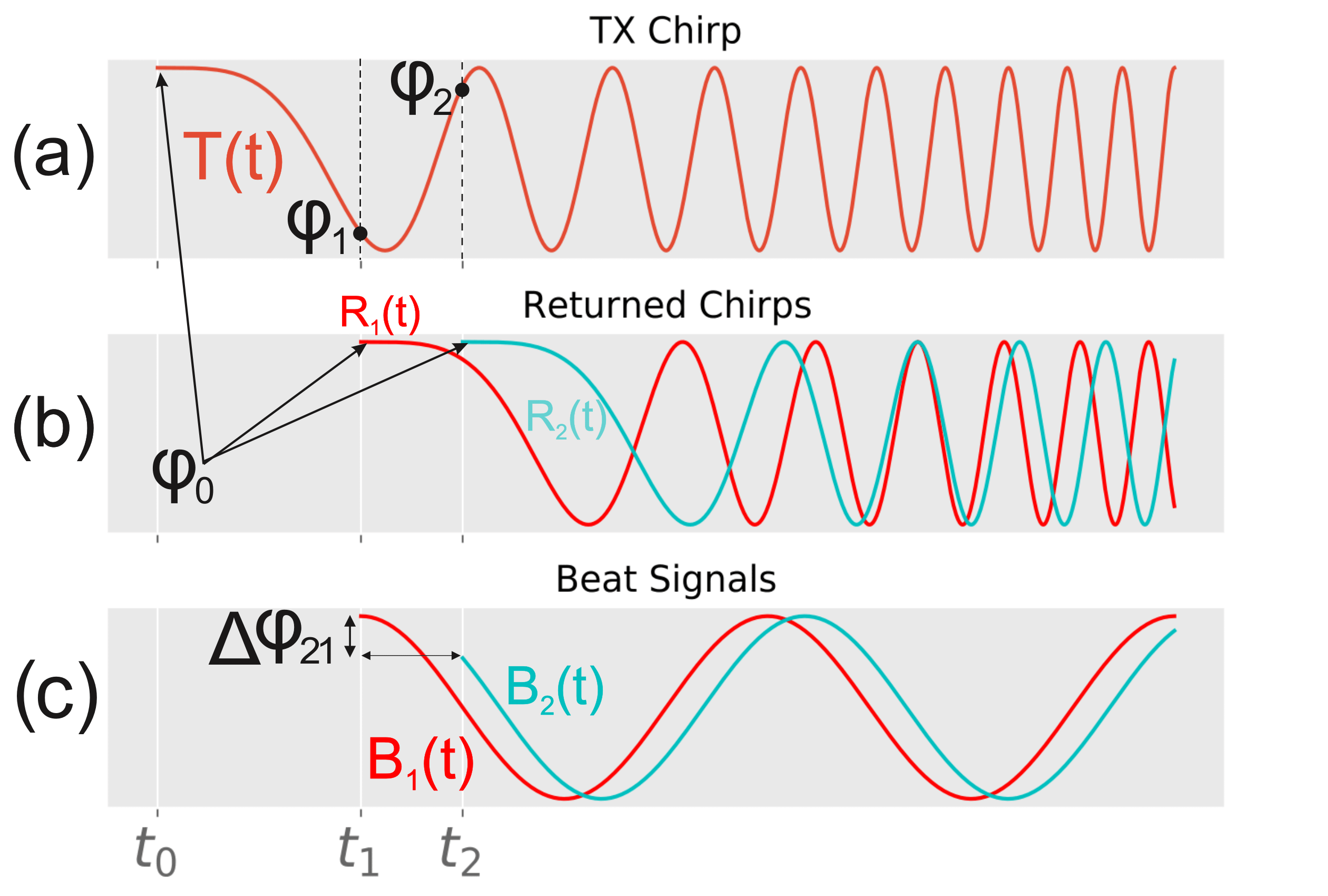}
\caption{Velocity measurement using FMCW radar: (a) Time representation of a chirp. The chirp is transmitted with an initial phase $\varphi _0$. (b) Two consecutive returned chirps $R_1(t)$, $R_2(t)$. (c) Two consecutive IF signals $B_1(t)$, $B_2(t)$ with the same beat frequency. The velocity of the object is proportional to the phase difference $\Delta\varphi _{21}$ between the phases of the IF signals.}
\label{fig:chirp tx and rx}
\end{figure}

To measure velocity, the radar utilizes pairs of chirps $T(t)$ that are transmitted consecutively, as depicted in Figure~\ref{fig:chirp tx and rx}(b). The calculation of the velocity relies on the fact that the change in the beat frequency between adjacent IF signals in the frame is negligible compared to the phase shift of the IF signal. This assumption holds since the phase of the mmWave signal is more sensitive to object movement, in contrast to the return delay, which is practically constant between the chirps in the same frame, due to the Doppler effect and the fast chirp modulation. Similarly to the range measurement, the IF signal $B_i(t)$ (Figure~\ref{fig:chirp tx and rx}(c)) corresponding to each returned chirp $R_i(t) = T(t-t_d)$ is sampled and further processed, this time in order to calculate its phase, again using range-FFT. The relative velocity $v$ of the moving object to the radar can be derived from the phase difference $\Delta\varphi_{i+1,i}$ between adjacent chirps' IF signals $B_{i+1},B_i$:\[v = \frac{\lambda\Delta\varphi_{i+1,i}}{4\pi T_{c}}\] where $T_{c}$ and $\lambda$ are the chirp duration and the radar signal wavelength, respectively (see~\cite{iovescu2017fundamentals}).

\section{The Adversarial FMCW Radar Model}
\begin{figure}[t]
\centering
\includegraphics[scale=0.12]{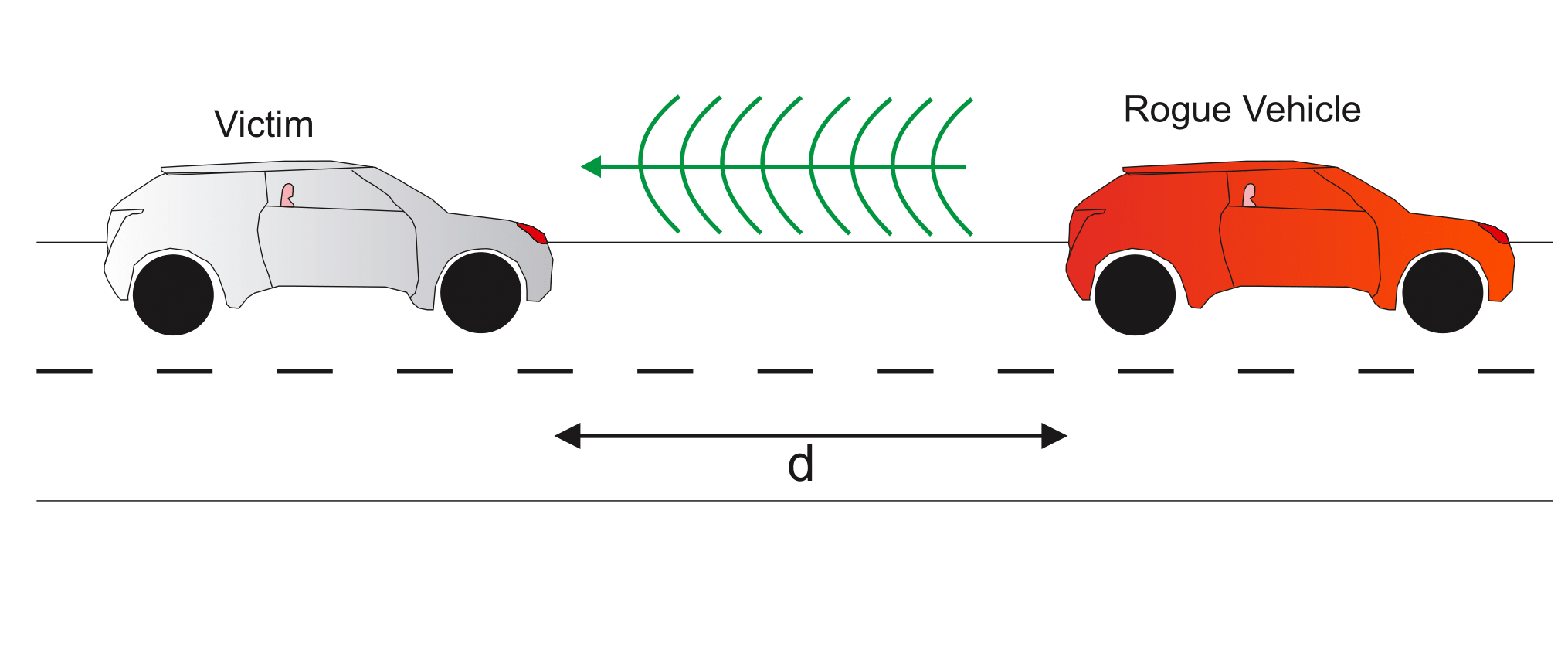}
\caption{The attack scenario. The victim vehicle is behind the attacker's vehicle, on the same traffic lane. The attacker sends a spoofing signal to the victim in order to perform the attack.}
\label{fig:Attack Scenario}
\end{figure}
\begin{figure*}[t]
\centering
\includegraphics[scale=0.25]{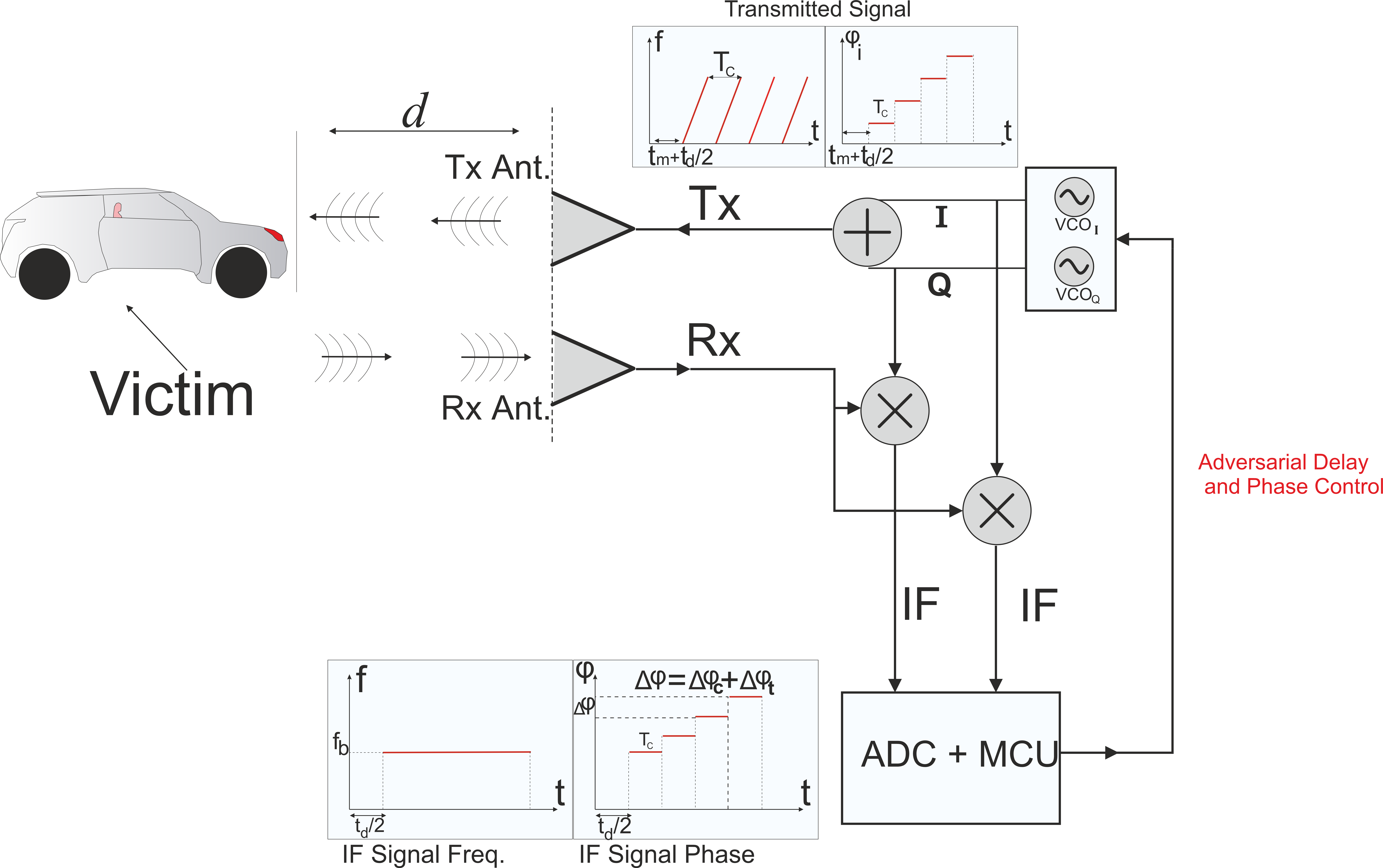}
\caption{ The Adversarial FMCW System: The original radar signal is received from the victim. Then, it is mixed with an internal synchronised chirp, to calculate the phase differences $\Delta\varphi_c$ and $\Delta\varphi_t$ between each consecutive pair of IF signals. The Adversary transmitter uses a quadrature modulator to control the transmitted phase and timing of each adversary chirp, to manipulate the victims measured velocity and range simultaneously. It sets the delay to $t_a$ and sets the phase to $\varphi_a$ compensating for $\Delta\varphi_c$ and $\Delta\varphi_t$ so the victim radar measures distance $\hat{d}$ and velocity $\hat{v}$.}
\label{fig:System Summary}
\end{figure*}

\subsection{Attack Model}
In our scenario the victim vehicle is behind the attacker's vehicle and has an FMCW radar installed, facing forward, as depicted in Figure~\ref{fig:Attack Scenario}. We assume the attacker has a modified radar system in his vehicle facing back. The attacker can send a much more powerful signal compared to the real reflected signal arriving at the victim. This assumption easily holds, even without extra amplification, as the attacker's signal only needs to traverse half of the distance, compared to the victim radar's signal. We assume that the victim radar parameters are known to the attacker. Moreover, the victim's true relative distance $d$ and relative velocity $v$ from the attacker are already measured by the attacker's radar, in order to be able to perform the attack.

\subsection{Attack System Architecture}The adversarial system is depicted in Figure~\ref{fig:System Summary}. The adversarial radar is a modified version of the victim radar: essentially the only change is the additional delay and phase control logic. Our system is a complex baseband (quadrature) radar that simultaneously controls the delay (to spoof range measurements) and phase (to spoof the velocity). 

At a high level, the adversarial radar starts by generating internal chirps with the same parameters as the victim, but without transmitting them. Once it receives the chirps transmitted by the victim radar, it mixes the first $n$ chirps out of the frame with its internal chirps to produce its own IF signal. The attack system utilizes this IF signal to synchronise to the victim, and tune the attack parameters. Once it is synchronised and tuned, the attack radar ignores the victim's remaining chirps in the frame, and instead actively transmits $N-n$ spoofed return signals. The synchronization and tuning techniques we use are explained below.

\subsection{Range Manipulation}
By precisely controlling the time the spoofed signal is received at the victim radar, the attacker can manipulate the victim radar to measure a delay $t_a$, which is different than the original delay $t_d$, leading to an error in the calculated distance. In order to spoof the victim radar to measure an arbitrary distance $\hat{d}$, the attacker should know the exact transmit time of the current frame of chirps. To achieve synchronisation with the victim's transmit time, the attacker estimates the arrival time to his Rx antenna of the first $n$ chirps in the frame. Since the first $n$ chirps are used in order to synchronise to the victim, only the remaining $N-n$ chirps in the frame are used for the range and velocity manipulation itself. Due to the fact that $N\gg n$, the chirps used for synchronisation have little effect on the manipulated range and velocity measured by the victim, since it is done by averaging the measurements of each frame, or by using two dimensional range-FFT~\cite{lutz2014fast}. The transmition time can be estimated by down-converting the arriving signal and estimating the TOA (Time of Arrival) using a relatively high speed ADC, or by using HCM (Half Chirp Modulation)~\cite{goodin2019predicting}. Considering that the chirp duration $T_c$, the true distance $d$ and also the true delay $t_d$ are known, the attacker knows when the next chirp will be transmitted, and thus can decide when to emit an adversarial set of chirps, which will modify the delay measured by the victim radar. The attacker can force the victim to measure a distance $\hat{d}$ corresponding to a specific adversary delay $t_a$, by delaying or preceding the returned signal the victim senses. Since the adversarial radar is actively transmitting, its spoofed signals will overpower the legitimate echos reflected from the adversarial vehicle. 
\subsection{Manipulating The Velocity}
As depicted in Figure~\ref{fig:chirp tx and rx}, in normal operation, the phase difference the victim radar measures between two consecutive chirps is given by:\[\Delta\varphi _{21}\ = \angle{B_2(t)} -\angle{B_1(t)}\]
\[=(\varphi _2 - \varphi _0) - (\varphi _1 - \varphi _0) = \varphi _2 -\varphi _1\]
where $\varphi_2$ and $\varphi_1$ are the phases $\angle{B_1(t)|_{t=t_1}}$ and $\angle{B_2(t)|_{t=t_2}}$ of two consecutive IF signals from two consecutive chirps, and $t_1$, $t_2$ are the delays corresponding to each chirp. 
As seen in Figure~\ref{fig:chirp tx and rx}(b), the victim FMCW radar expects to receive a delayed cloned version of the signal it sends, i.e., with the same phase $\varphi _0$, when it hits the Rx antenna. The method for velocity manipulation we propose takes advantage of this fact. 
To manipulate the velocity, the attacker transmits a frame of chirps with \emph{changing} phases. The difference $\Delta\varphi_{21}$ is known to the attacker because the true velocity of the victim is known. Therefore, by changing the phase $\varphi _0$ of the spoofed return signal separately for each chirp, it is possible for the attacker to control the velocity the victim measures. This can be done regardless of the victim radar's current transmitting chirp phase: the victim's measured phase difference depends only on the phase difference between consecutive attacker-generated returned chirps, and does not depend on the victim radar itself.

\emph{Controlling The Phase:}
As demonstrated in Figure~\ref{fig:System Summary}, the attacker should send a sequence of chirps with changing phases. To do so, the attacker utilizes the quadrature manner of the complex baseband radar in order to control the phase of the transmitted chirp, similarly to PSK (Phase Shift Keying) techniques. In order to make the victim radar measure a specific velocity $\hat{v}$,  
the phase of each chirp of the signal transmitted by the attacker should change with each consecutive chirp. The manipulated phase between consecutive chirps $\Delta\varphi _{i+1,i}$ is given by: \[\Delta\varphi _{i+1,i} = \frac{4\pi \hat{v}T_{c}}{\lambda}\]

\section{The Process of Attacking a Victim Radar}
\subsection{Range Manipulation}
Beyond Figure~\ref{fig:System Summary}, Figure~\ref{fig:vel-spoof} shows the evolution of the attack signal. First, the adversarial radar enters a ``waiting for trigger'' mode. When it receives the first chirp, it synchronises with the victim radar using estimation of TOA of the signal, or using an internal HCM. This synchronisation  occurs only once in the whole process of position and velocity manipulation. As seen in Figure~\ref{fig:System Summary}, when the attacker is synchronised to the victim, it can generate an IF signal $B_i(t)$ by mixing an internal synchronised version of the victim radar chirp with the incoming signal. This IF signal is depicted in Figure~\ref{fig:vel-spoof}(a).
Now, it can manipulate the range the victim radar measures by precisely controlling the adversarial delay $t_a = t_d + t_m$, where $t_d$ is the actual time of flight (TOF) and $t_m$ is the manipulated delay the attacker adds. Since the legitimate round trip is $t_d$, the attacker should transmit the spoofing signal at $t_d/2 + t_m$, taking into account that the spoofing signal should still traverse the return path, which adds a delay of $t_d/2$. This is depicted in Figure~\ref{fig:vel-spoof}(c).

\subsection{Velocity Manipulation}
\begin{figure*}[t]
\centering
\includegraphics[scale=0.4]{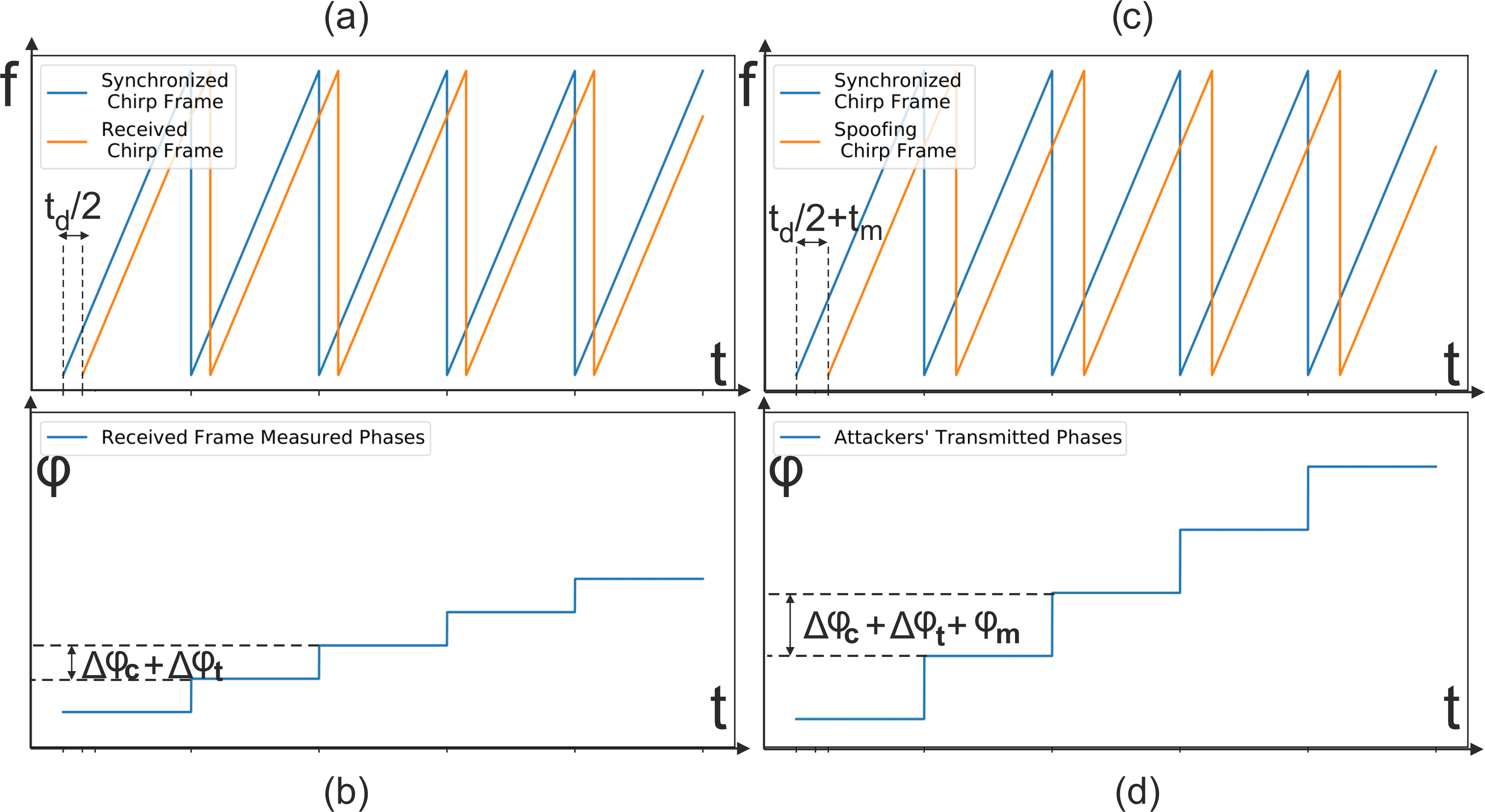}
\caption{The Process of Range and Velocity Spoofing: (a) The received chirp at the adversary radar (orange) and the synchronised internal chirp (blue). (b) The changing phase between each internal measured IF signal in the adversary radar. $\Delta\varphi_c$ is the phase accumulated because of frequency difference from the victim, and $\Delta\varphi_t$ is the phase change caused by the relative velocity. (c) The transmitted adversary signal spectrogram, with the controlled delay $t_a=t_d/2+t_m$ corresponding to the spoofed distance $\hat{d}$. (d) The initial phase of each adversarial chirp emitted from the attacker as function of time. The phase is composed of the compensation for the phases $\Delta\varphi_c$ and $\Delta\varphi_t$, and the controlled phase $\varphi_m$ corresponding to the spoofed velocity $\hat{v}$.}
\label{fig:vel-spoof}
\end{figure*}

Although the victim radar and the adversarial radar are both using the same central frequency and sweep bandwidth, there is some difference between them, because of the TCXO (Temperature Controlled Crystal Oscillator) frequency tolerance. Therefore, there is an accumulated phase $\Delta\varphi_c$ with each internal IF signal $B_i(t-nT_c)$ in the attackers' radar. Furthermore, there is another phase accumulation $\Delta\varphi _t$, due to the true velocity difference $v$ between the victim and the adversary. The effect of the phases $\Delta\varphi_c$ and $\Delta\varphi _t$ on the internal IF signals $B_i(t-nT_c)$ is depicted in Figure~\ref{fig:vel-spoof} (b). The attacker takes these phases into account, and compensates them when manipulating the velocity.
The total adversarial phase increment $\Delta\varphi_a = \Delta\varphi_c + \Delta\varphi_t + \varphi_m$ of each chirp the attacker sends back consists the compensation of the phases $\Delta\varphi_c$ and $\Delta\varphi_t$, and the manipulating phase $\varphi_m$. The manipulating phase $\varphi_m$, is the phase determining the current measurement change in relative velocity the victim will measure. By changing this phase with each chirp that the adversary transmits, it can precisely control the velocity $\hat{v}$ measured  by the victim, as described in Figure~\ref{fig:vel-spoof}(d).
Velocity and distance spoofing can occur independently, as the phase and the timing of the returned chirps are not related to each other. Therefore, the adversarial radar can manipulate both of them simultaneously. The adversary signal $S_{Adv}$ transmitted by the attacker in order to manipulate both range and velocity is the concatenation (sum) of the $N-n$ manipulated chirps:
\[S_{Adv} = \sum_{i=n+1}^{N}T(t-(\frac{t_d}{2} + t_{m} + i\cdot T_c))\cdot e^{j\cdot i\Delta\varphi_a}\]
Where $T(t)$ is the chirp transmitted by the victim radar, and $i$ is an index iterating over the $N-n$ manipulated chirps.

\begin{figure}[t]
\centering
\includegraphics[scale=0.3]{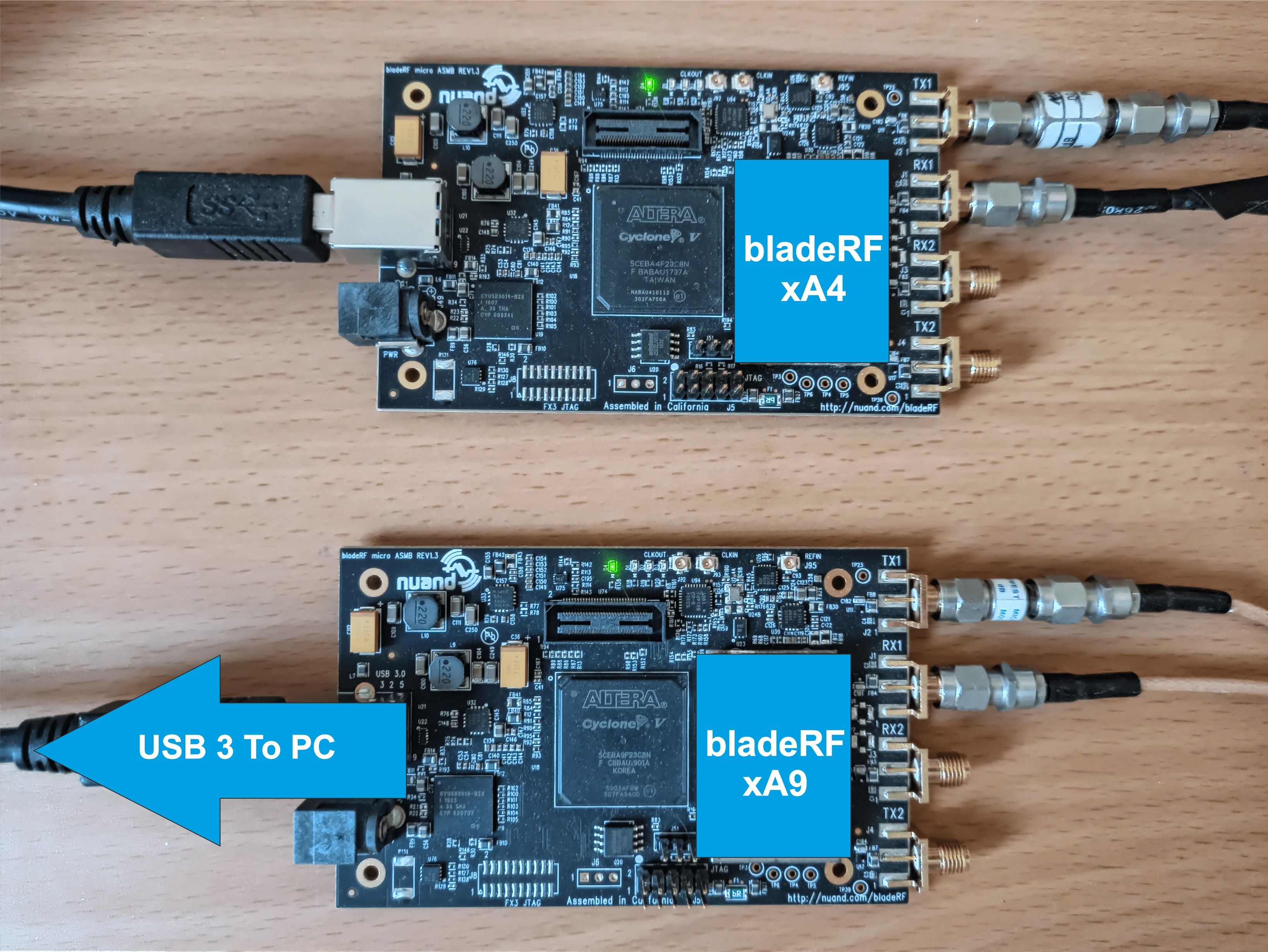}
\caption{The attack radar (xA9) below and the victim radar (xA4) above. The two radars are connected by 15m RF cables beyond the photo's right margin. }
\label{fig:system_photo}
\end{figure}

\section{The Proof of Concept Setup}
The setup we implemented (see Figure~\ref{fig:system_photo}) consists of two Software Defined Radios (SDR): one is the attack radar, and the other represents the victim. Both SDRs are models of bladeRF 2.0 micro. 
The bladeRF 2.0 micro is a 2x2 MIMO, 47MHz to 6 GHz frequency range, USB 3.0 Software Defined Radio~\cite{nuand20163}. It provides a powerful waveform development platform.
We used the xA9 model to build the attack radar, and for the victim FMCW radar we used the xA4 model. The libbladeRF open source library was used for communicating with the devices, from a Windows machine.
The interface we selected is the bladeRF Synchronous Interface, which allows transmitting bursts of samples at a specified timestamp. The timestamp is a free-running counter in the FPGA, incremented at the sample rate defined, with each outgoing sample.
For the ease of prototyping, the victim Tx and Rx were connected the attacker's Rx and Tx respectively by 15m-long  RF cables. 
To simulate greater and more realistic distances between vehicles with these cables, a one-time calibration process was carried out. In this process, the victim radar transmitted a set of chirps and the attack radar immediately returned the signal it received. Then, a fixed delay was added to the response time of the attacker, so the victim radar measured a distance $d = 60m$. Due to the fact that the length of the cable imitating the distance was fixed, the relative velocity $v$ simulated between the victim radar and the attacker was fixed to zero.

To add an arbitrary phase shift $\Delta\varphi$ to the transmitted spoofed signal, we used the quadrature modulator of the attacking SDR. The attacker controls the transmitted chirps' phases by utilizing Euler's formula
\[
A(t)\cdot e^{j\cdot\Delta\varphi} = A(t)\cdot\cos\Delta\varphi + jA(t)\cdot\sin\Delta\varphi
\]
and multiplying the in-phase and quadrature-phase parts of the modulator with the controlled amplitudes.

 In order to synchronise to the victim, we estimated the TOA, by sensing the first sample of the IF signal which is above an arbitrary threshold of noise.
The center frequency $f_c$ used in the setup was 1GHz, the full bandwidth of the radar was 28MHz, the chirp duration $T_c$ was set to 1ms, and $n$ was set to 2 chirps.


\begin{figure*}[t]
\centering
\includegraphics[scale=0.2]{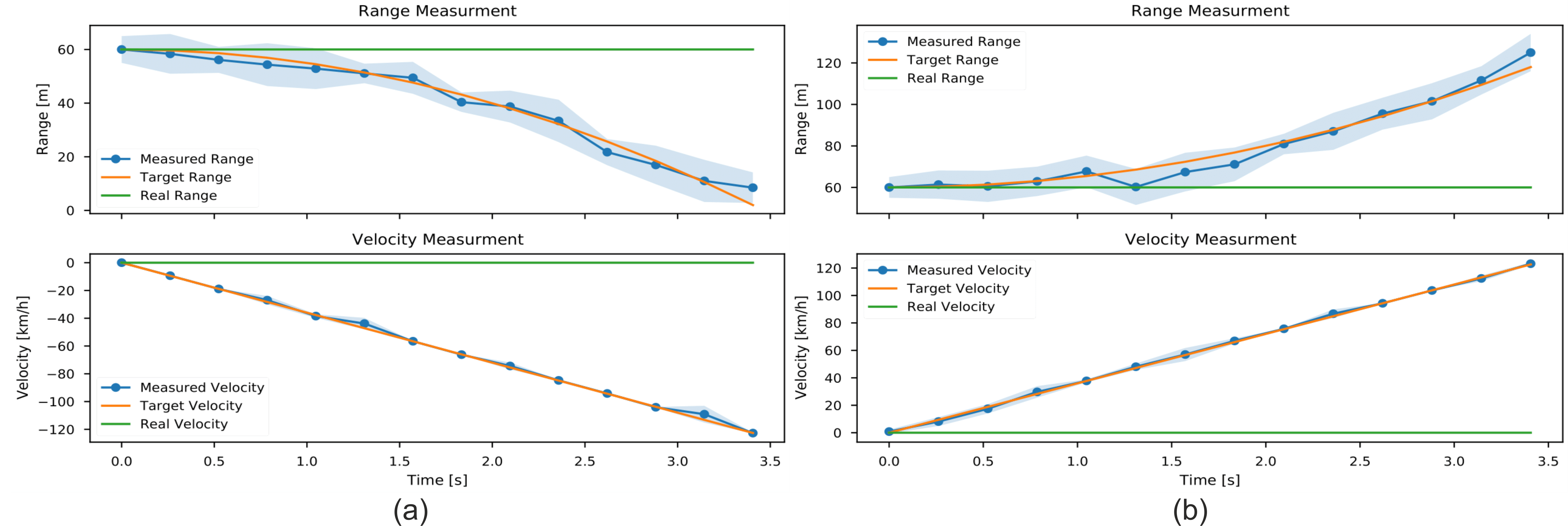}
\caption{Results of Simultaneous Spoofing of Distance and Velocity: The green line shows the true range and velocity. The orange curve shows the desired manipulated range and velocity and the blue dots show the average values measured by the victim. The time between the measurements is 250ms. The shaded area shows the errors (+/- one standard deviation). (a) Spoofing an emergency brake. (b) Spoofing a scenario where an obstacle appears to be moving away from the victim radar.}
\label{fig:sim_vel_pos}
\end{figure*}

\section{Results and Discussion}
We simulated two scenarios: a sudden phantom stop and a sudden phantom acceleration. Each scenario was simulated 15 times. Simultaneous spoofing of distance and velocity is demonstrated in Figure~\ref{fig:sim_vel_pos}. 

Figure~\ref{fig:sim_vel_pos}(a) describes a scenario where the attacking radar simultaneously spoofs the distance and velocity representing a phantom emergency break in the adversary's vehicle in front of it: the victim's measured relative velocity $\hat{v}$ drops at a constant deceleration of $-10~m/sec^2$ spoofing a loss of $120~km/h$, and the measured relative distance drops at a quadratic rate from $60m$ to $0$ over about 3.5 sec: This rate of deceleration represents an emergency brake with ABS assistance \cite{kudarauskas2007analysis}. Note that the errors are in the order of the bandwidth resolution limit of FMCW, and only limited by the SDR relatively low bandwidth capabilities. Such a scenario will probably trigger an emergency break by the victim.

Figure~\ref{fig:sim_vel_pos}(b) demonstrates another security hazard: the vehicle in front of the victim is made to appear accelerating, when it actually remains at a fixed distance $d = 60m$ and a fixed relative velocity $v = 0$. In such a case, the victim car might decide it can accelerate, leading to an accident and injuries to the driver and passengers inside of the car. The figure shows that the victim measures a velocity that mimics a fixed acceleration of $10~m/sec^2$, and the range measurements increase quadratically, matching the desired curve.

Note that in both scenarios the desired spoofed orange curves are within the measured error range, and in particular the measured velocities are very close to the intended values.

These attacks are very difficult for detection, since they are indistinct from a real scenario from the perspective of the victim vehicle. They cannot be mitigated by intra-radar sensor-fusion, since the range and velocity measurements are coherent with each other and consistent with the laws of physics.

\section{Countermeasures}
Our experiments showed the feasibility of simultaneous spoofing of the distance and velocity, and the security threats to the FMCW system.
There are several countermeasures that can be used:
\subsubsection{Sensor fusion with LiDAR/camera}
Since modern vehicles are equipped with several range and velocity sensors, sensor fusion with non-radar sensors may mitigate the attack. I.e., when the mmWave sensor warns against impending collision, other sensors should agree. For example, in~\cite{yang2021secure}, the authors are proposing a sensor fusion framework that uses multiple sensors for measuring the same physical variable to create redundancy. Then it is used to estimate the real measurements and mitigate a physical level attack on sensor level. However, there are scenarios, such as bad weather conditions, where the only reliable sensor in the car is the mmWave radar. In such cases, sensor fusion can mislead the vehicle to believe that it has no danger ahead, while it is actually about to collide with the car in front of it, or, conversely, to halt when no emergency exists.

\subsubsection{Phase Randomization} The velocity spoofing attack relies on the fact that the victim radar is expecting the returned signal to arrive with the same phase it was transmitted at. To break this assumption, one can send chirp signals with a randomized phases, which are known to the victim radar. Since the phases are randomized, the adversarial radar can't compensate the phase difference. This approach requires more complex circuity, e.g., another quadrature mixer, and therefore increases the cost of the radar. Moreover, sending another phase with each chirp adds another phase error to an environment that is already very error-prone.

\subsubsection{Frequency Randomization} Another possible method to countermeasure the proposed attacks is to use random frequency hoping. This approach will make the attack radar measure a high beat frequency when implementing the proposed attack, and therefore will make it hard to synchronise to the victim radar. The BlueFMCW for example is a novel frequency hoping radar, that uses frequency randomization in order to countermeasure attacks~\cite{moon2020bluefmcw}. In~\cite{nashimoto2021low}, the authors also propose a random-chirp modulation based on randomization of frequency. Both mitigation systems require additional hardware in order to cope with the attack.

The authors of~\cite{moon2020bluefmcw,nashimoto2021low} provide only numeric simulation level results for the countermeasures, and still require further work on a real equipment in order to convince their effectiveness. Furthermore, even in non-adversarial environments, the large number of vehicles on the same road already leads to frequency spectrum density. Therefore, frequent changes in frequency may help not only countermeasure the described attacks, but can also offer interference mitigation~\cite{bourdoux2017phenomenology}.

\subsubsection{RSSI Measurements} Since the attacker assumes it will overpower the legitimate echos reflected from his vehicle with his adversarial signal, the victim can use the Received Signal Strength Indicator (RSSI) to countermeasure the attack. For instance, the victim should look for a pattern of $n$ weak chirps out of the total $N$ chirps in the frame. This approach may fail, when there is a legitimate change in the power of the signal received, due to weather conditions or a fading channel.  

\section{Conclusions}

The security of the ADAS and autonomous vehicles sensors in the face of cyber attacks is crucial for the future of the automotive industry. In this paper, we proposed a system to attack an automotive FMCW mmWave radar, that uses fast chirp modulation. Our attack system is capable of spoofing the distance and velocity measured by the victim vehicle simultaneously, presenting phantom measurements coherent with the laws of physics governing vehicle motion. The attacking radar controls the delay in order to spoof its distance, and uses phase compensation and control in order to spoof its velocity.
We  demonstrated the spoofing attack by building a proof-of-concept hardware-based system, using a Software Defined Radio.
We successfully demonstrated two real world scenarios, in which the victim radar is spoofed to detect either a phantom emergency stop or a phantom acceleration, measuring coherent range and velocity. 
We believe that with the proliferation of vehicular FMCW radars, the demonstrated attack system could pose a threat to AV and ADAS safety, and we propose several possible mitigations.

\bibliographystyle{IEEEtranS}
\bibliography{references}

\begin{thebibliography}{10}
\providecommand{\url}[1]{#1}
\csname url@samestyle\endcsname
\providecommand{\newblock}{\relax}
\providecommand{\bibinfo}[2]{#2}
\providecommand{\BIBentrySTDinterwordspacing}{\spaceskip=0pt\relax}
\providecommand{\BIBentryALTinterwordstretchfactor}{4}
\providecommand{\BIBentryALTinterwordspacing}{\spaceskip=\fontdimen2\font plus
\BIBentryALTinterwordstretchfactor\fontdimen3\font minus
  \fontdimen4\font\relax}
\providecommand{\BIBforeignlanguage}[2]{{%
\expandafter\ifx\csname l@#1\endcsname\relax
\typeout{** WARNING: IEEEtranS.bst: No hyphenation pattern has been}%
\typeout{** loaded for the language `#1'. Using the pattern for}%
\typeout{** the default language instead.}%
\else
\language=\csname l@#1\endcsname
\fi
#2}}
\providecommand{\BIBdecl}{\relax}
\BIBdecl

\bibitem{bourdoux2017phenomenology}
A.~Bourdoux, K.~Parashar, and M.~Bauduin, ``Phenomenology of mutual
  interference of {FMCW} and {PMCW} automotive radars,'' in \emph{2017 IEEE
  Radar Conference (RadarConf)}.\hskip 1em plus 0.5em minus 0.4em\relax IEEE,
  2017, pp. 1709--1714.

\bibitem{cao2019adversarial}
Y.~Cao, C.~Xiao, B.~Cyr, Y.~Zhou, W.~Park, S.~Rampazzi, Q.~A. Chen, K.~Fu, and
  Z.~M. Mao, ``Adversarial sensor attack on {LiDAR}-based perception in
  autonomous driving,'' in \emph{Proceedings of the 2019 ACM SIGSAC Conference
  on Computer and Communications Security}, 2019, pp. 2267--2281.

\bibitem{changalvala2019lidar}
R.~Changalvala and H.~Malik, ``{LiDAR} data integrity verification for
  autonomous vehicle,'' \emph{IEEE Access}, vol.~7, pp. 138\,018--138\,031,
  2019.

\bibitem{eykholt2018robust}
K.~Eykholt, I.~Evtimov, E.~Fernandes, B.~Li, A.~Rahmati, C.~Xiao, A.~Prakash,
  T.~Kohno, and D.~Song, ``Robust physical-world attacks on deep learning
  visual classification,'' in \emph{Proceedings of the IEEE Conference on
  Computer Vision and Pattern Recognition}, 2018, pp. 1625--1634.

\bibitem{fung2017sensor}
M.~L. Fung, M.~Z. Chen, and Y.~H. Chen, ``Sensor fusion: A review of methods
  and applications,'' in \emph{2017 29th Chinese Control And Decision
  Conference (CCDC)}.\hskip 1em plus 0.5em minus 0.4em\relax IEEE, 2017, pp.
  3853--3860.

\bibitem{gao2019experiments}
X.~Gao, G.~Xing, S.~Roy, and H.~Liu, ``Experiments with mmwave automotive radar
  test-bed,'' in \emph{2019 53rd Asilomar Conference on Signals, Systems, and
  Computers}.\hskip 1em plus 0.5em minus 0.4em\relax IEEE, 2019, pp. 1--6.

\bibitem{goodin2019predicting}
C.~Goodin, D.~Carruth, M.~Doude, and C.~Hudson, ``Predicting the influence of
  rain on {LiDAR} in {ADAS},'' \emph{Electronics}, vol.~8, no.~1, p.~89, 2019.

\bibitem{iovescu2017fundamentals}
\BIBentryALTinterwordspacing
C.~Iovescu and S.~Rao, ``The fundamentals of millimeter wave sensors,''
  \emph{Texas Instruments}, pp. 1--8, 2017. [Online]. Available:
  \url{https://www.mouser.ee/pdfdocs/mmwavewhitepaper.pdf}
\BIBentrySTDinterwordspacing

\bibitem{jagielski2018threat}
M.~Jagielski, N.~Jones, C.-W. Lin, C.~Nita-Rotaru, and S.~Shiraishi, ``Threat
  detection for collaborative adaptive cruise control in connected cars,'' in
  \emph{Proceedings of the 11th ACM Conference on Security \& Privacy in
  Wireless and Mobile Networks}, 2018, pp. 184--189.

\bibitem{kapoor2018detecting}
P.~Kapoor, A.~Vora, and K.-D. Kang, ``Detecting and mitigating spoofing attack
  against an automotive radar,'' in \emph{2018 IEEE 88th Vehicular Technology
  Conference (VTC-Fall)}.\hskip 1em plus 0.5em minus 0.4em\relax IEEE, 2018,
  pp. 1--6.

\bibitem{komissarov2019partially}
R.~Komissarov, V.~Kozlov, D.~Filonov, and P.~Ginzburg, ``Partially coherent
  radar unties range resolution from bandwidth limitations,'' \emph{Nature
  communications}, vol.~10, no.~1, pp. 1--9, 2019.

\bibitem{kudarauskas2007analysis}
N.~Kudarauskas, ``Analysis of emergency braking of a vehicle,''
  \emph{Transport}, vol.~22, no.~3, pp. 154--159, 2007.

\bibitem{lin2016design}
J.-J. Lin, Y.-P. Li, W.-C. Hsu, and T.-S. Lee, ``Design of an {FMCW} radar
  baseband signal processing system for automotive application,''
  \emph{SpringerPlus}, vol.~5, no.~1, p.~42, 2016.

\bibitem{lutz2014fast}
S.~Lutz, D.~Ellenrieder, T.~Walter, and R.~Weigel, ``On fast chirp modulations
  and compressed sensing for automotive radar applications,'' in \emph{2014
  15th International Radar Symposium (IRS)}.\hskip 1em plus 0.5em minus
  0.4em\relax IEEE, 2014, pp. 1--6.

\bibitem{mitomo201077}
T.~Mitomo, N.~Ono, H.~Hoshino, Y.~Yoshihara, O.~Watanabe, and I.~Seto, ``A 77
  {GHz} 90 nm {CMOS} transceiver for {FMCW} radar applications,'' \emph{IEEE
  journal of solid-state circuits}, vol.~45, no.~4, pp. 928--937, 2010.

\bibitem{miura2019low}
N.~Miura, T.~Machida, K.~Matsuda, M.~Nagata, S.~Nashimoto, and D.~Suzuki, ``A
  low-cost replica-based distance-spoofing attack on mmwave {FMCW} radar,'' in
  \emph{Proceedings of the 3rd ACM Workshop on Attacks and Solutions in
  Hardware Security Workshop}, 2019, pp. 95--100.

\bibitem{moon2020bluefmcw}
T.~Moon, J.~Park, and S.~Kim, ``{BlueFMCW}: Random frequency hopping radar for
  mitigation of interference and spoofing,'' \emph{arXiv preprint
  arXiv:2008.00624}, 2020.

\bibitem{nashimoto2021low}
S.~Nashimoto, D.~Suzuki, N.~Miura, T.~Machida, K.~Matsuda, and M.~Nagata,
  ``Low-cost distance-spoofing attack on {FMCW} radar and its feasibility study
  on countermeasure,'' \emph{Journal of Cryptographic Engineering}, pp. 1--10,
  2021.

\bibitem{nuand20163}
{Nuand bladeRF, USB}, ``3.0 software defined radio manual,'' 2016.

\bibitem{ranganathan2012physical}
A.~Ranganathan, B.~Danev, A.~Francillon, and S.~Capkun, ``Physical-layer
  attacks on chirp-based ranging systems,'' in \emph{Proceedings of the fifth
  ACM conference on Security and Privacy in Wireless and Mobile Networks},
  2012, pp. 15--26.

\bibitem{song2018physical}
D.~Song, K.~Eykholt, I.~Evtimov, E.~Fernandes, B.~Li, A.~Rahmati, F.~Tramer,
  A.~Prakash, and T.~Kohno, ``Physical adversarial examples for object
  detectors,'' in \emph{12th {USENIX} Workshop on Offensive Technologies
  (WOOT'18)}, 2018.

\bibitem{stove1992linear}
A.~G. Stove, ``Linear {FMCW} radar techniques,'' in \emph{IEE Proceedings F
  (Radar and Signal Processing)}, vol. 139, no.~5.\hskip 1em plus 0.5em minus
  0.4em\relax IET, 1992, pp. 343--350.

\bibitem{sun2020control}
Z.~Sun, S.~Balakrishnan, L.~Su, A.~Bhuyan, P.~Wang, and C.~Qiao, ``Who is in
  control? practical physical layer attack and defense for {mmWave} based
  sensing in autonomous vehicles,'' \emph{arXiv preprint arXiv:2011.10947},
  2020.

\bibitem{thing2016autonomous}
V.~L. Thing and J.~Wu, ``Autonomous vehicle security: A taxonomy of attacks and
  defences,'' in \emph{2016 IEEE International Conference on Internet of Things
  (iThings) and IEEE Green Computing and Communications (GreenCom) and IEEE
  Cyber, Physical and Social Computing (CPSCom) and IEEE Smart Data
  (SmartData)}.\hskip 1em plus 0.5em minus 0.4em\relax IEEE, 2016, pp.
  164--170.

\bibitem{tong2015fast}
Z.~Tong, R.~Renter, and M.~Fujimoto, ``Fast chirp {FMCW} radar in automotive
  applications,'' in \emph{IET International Radar Conference 2015}.\hskip 1em
  plus 0.5em minus 0.4em\relax IET, 2015, pp. 1--4.

\bibitem{wang2019multi}
Z.~Wang, Y.~Wu, and Q.~Niu, ``Multi-sensor fusion in automated driving: A
  survey,'' \emph{IEEE Access}, vol.~8, pp. 2847--2868, 2019.

\bibitem{winkler2007range}
V.~Winkler, ``Range {Doppler} detection for automotive {FMCW} radars,'' in
  \emph{2007 European Radar Conference}.\hskip 1em plus 0.5em minus 0.4em\relax
  IEEE, 2007, pp. 166--169.

\bibitem{yan2016can}
C.~Yan, W.~Xu, and J.~Liu, ``Can you trust autonomous vehicles: Contactless
  attacks against sensors of self-driving vehicle,'' \emph{DEF CON}, vol.~24,
  no.~8, p. 109, 2016.

\bibitem{yang2021secure}
T.~Yang and C.~Lv, ``A secure sensor fusion framework for connected and
  automated vehicles under sensor attacks,'' \emph{arXiv preprint
  arXiv:2103.00883}, 2021.

\bibitem{ziebinski2016survey}
A.~Ziebinski, R.~Cupek, H.~Erdogan, and S.~Waechter, ``A survey of {ADAS}
  technologies for the future perspective of sensor fusion,'' in
  \emph{International Conference on Computational Collective
  Intelligence}.\hskip 1em plus 0.5em minus 0.4em\relax Springer, 2016, pp.
  135--146.

\end{thebibliography}


\end{document}